\documentclass[conference]{IEEEtran}
\IEEEoverridecommandlockouts
\usepackage{cite}
\usepackage{amsmath,amssymb,amsfonts}
\usepackage{algorithmic}
\usepackage{graphicx}
\usepackage{textcomp}
\usepackage{xcolor}
\usepackage{booktabs}
\usepackage{lscape}
\usepackage{longtable}
\usepackage{array}
\usepackage [english]{babel}
\usepackage [autostyle, english = american]{csquotes}
\MakeOuterQuote{"}
\def\BibTeX{{\rm B\kern-.05em{\sc i\kern-.025em b}\kern-.08em
    T\kern-.1667em\lower.7ex\hbox{E}\kern-.125emX}}
\begin{document}

\title{Explainable Artificial Intelligence and Cybersecurity: A Systematic Literature Review}

\author{\IEEEauthorblockN{1\textsuperscript{st} Carlos Frederico D'Almeida e Mendes}
\IEEEauthorblockA{\textit{Institute of Computing} \\
\textit{Federal University of Bahia}\\
Salvador, Brazil \\
carlosfam@ufba.br}
\and
\IEEEauthorblockN{2\textsuperscript{nd} Tatiane Nogueira Rios}
\IEEEauthorblockA{\textit{Institute of Computing} \\
\textit{Federal University of Bahia}\\
Salvador, Brazil \\
tatiane.nogueira@ufba.br}
}

\maketitle

\begin{abstract}
Cybersecurity vendors consistently apply AI (Artificial Intelligence) to their solutions and many cybersecurity domains can benefit from AI technology. However, black-box AI techniques present some difficulties in comprehension and adoption by its operators, given that their decisions are not always humanly understandable (as is usually the case with deep neural networks, for example). Since it aims to make the operation of AI algorithms more interpretable for its users and developers, XAI (eXplainable Artificial Intelligence) can be used to address this issue. Through a systematic literature review, this work seeks to investigate the current research scenario on XAI applied to cybersecurity, aiming to discover which XAI techniques have been applied in cybersecurity, and which areas of cybersecurity have already benefited from this technology.
\end{abstract}


\section{Introduction}
A \emph{cyber risk} can be roughly defined as a factor that can lead to unwanted digital information leakage, sequestration, or destruction; unauthorized use of computer resources; unavailability of computer systems; or that compromises the integrity or confidentiality of digital data in general. Further discussion of this concept can be seen in \cite{refsdal2015cyber}.

The materialization of cyber risk can generate losses that, depending on their severity, can be treated as a \emph{cyber-incident}. In the business environment, the 5 consequences of a cyber-incident identified as the most negative are \cite{TrendMicro2021}:

\begin{itemize}
\item Customer turnover
\item Lost intellectual property (including trade secrets)
\item Disruption or damages to critical infrastructure
\item Cost of outside consultants and experts
\item Lost revenues
\end{itemize}

Cyber incidents are touted as the biggest risk factor for business in 2022, according to research conducted by insurer Allianz, even ahead of \textit{business interruption}, \textit{natural catastrophes}, and \textit{pandemic outbreak}\cite{Allianz2022}. 

Cybersecurity spans incredibly diverse specialties. The (ISC)² (International Information System Security Certification Consortium), maintainer of the CISSP(Certified Information Systems Security Professional) qualification, projects its exam in 8 domains:

\begin{itemize}
\item Security and Risk Management
\item Asset Security
\item Security Architecture and Engineering
\item Communications and Network Security
\item Identity and Access Management
\item Security Assessment and Testing
\item Security Operations
\item Software Development Security
\end{itemize}

This breadth of knowledge required is one of the factors why the cybersecurity industry is currently facing a shortage of qualified professionals. The worldwide workforce gap is estimated at 2.7 million professionals \cite{ISC2021}.

To reduce the industry's dependence on these sought-after analysts, cybersecurity vendors make extensive use of AI (Artificial Intelligence) in a variety of products. Some of the technologies already consolidated or in adoption by the market can be verified in \cite{GartnerEPP2021,GartnerCloud2021,GartnerNW2021,GartnerSO2021}, each of them using AI techniques to a greater or lesser extent.

According to technological research and consulting firm, Gartner \cite{Gartner2021}, there are 19 current prominent AI use cases that are directly relevant to the security and risk management leaders, some of which are:

\begin{itemize}
    \item Transaction Fraud Detection
    \item File-Based Malware Detection
    \item Process Behavior Analysis
    \item Abnormal System Behavior Detection
    \item Account Takeover Identification
    \item Asset Inventory and Dependency Mapping Optimization
    \item Web Domain and Reputation Assessment
\end{itemize}

Unfortunately, there are AI techniques whose operation is not transparent to the user and which also do not provide explanations on how they arrived at the generated result, as is usually the case with neural networks, for example. These are called "black box" AI techniques. It turns out that a better understanding by technology operators is desirable as it allows greater\cite{molnar2022}:

\begin{itemize}
    \item Trust in its decisions
    \item Social acceptance
    \item Ease of debugging and auditing
    \item Fairness (by the ease of bias detection)
    \item Assessment on the relevance of learned features
\end{itemize}

With this problem in mind, the concept of XAI (eXplainable Artificial Intelligence) was derived, whose goal is to make the operation of an AI algorithm more understandable for its users and developers. XAI can be defined as the set of AI methods capable of conveying to a suitably specialized observer how they arrived at a classification, regression, or prediction. The discussion about what constitutes ``understanding'' is heated. There is no peaceful and widely accepted definition, but attempts of formalization have already been well developed since at least \cite{Turing1950}. For this discussion, the reading of \cite{Aaronson2011} may be relevant.

Some features are particularly desirable in an XAI application \cite{Hansen2019}:
\begin{itemize}
    \item Understandability
    \item Fidelity: reasonable representation of what the AI system actually does.
    \item Sufficiency: detailed enough to justify the AI decision.
    \item Low Construction Overhead: not dominate the cost of designing the AI.
    \item Efficiency: not slow down the AI significantly.
\end{itemize}

With regard specifically to the application of XAI to cybersecurity, \cite{Vigano2018} and \cite{Paredes2021} address this matter in a high-level way, proposing a so-called \emph{desiderata} for the area and general architecture that can serve as a roadmap for guiding research efforts towards the development of XAI-based cybersecurity systems.

One way XAI algorithms can be classified is on whether interpretability is achieved by restricting the complexity of the machine learning model (\emph{intrinsic}) or by applying methods that analyze the model after training (\emph{post hoc}). Furthermore, depending on the scope of interpretability, they can be classified as \emph{global} (explain the entire model behavior) or \emph{local} (explain an individual prediction) \cite{molnar2022}.

Through a Systematic Literature Review (SLR), this work seeks to investigate the current research scenario on XAI applied to cybersecurity. 

The SLR follows 3 well-defined steps: search (query), analysis (quantitative and qualitative insights), and conclusion (response to the Main Research Question).
In the following sections, each of the SLR phases will be presented, with their development and results.

\section{SLR Phase 1: Search}

At this stage of the SLR, we defined the Main Research Question (MRQ), a set of Secondary Questions (SQ), the repositories to be searched, the language of the articles to be evaluated, the keywords and search query, and the inclusion and exclusion criteria for returned articles.

In order to investigate the current research scenario on XAI techniques applied to cybersecurity, the Main Research Question we aim to answer is:

\begin{center}
\emph{What are the XAI techniques used to promote more interpretable automated cyber risk classification?}
\end{center}

In addition, some secondary questions were elaborated, seeking to give more details to the research scenario in the area:
\begin{enumerate}
\item Which countries do most research on the subject come from?
\item What is the frequency of published studies on the subject?
\item How are studies in the area divided by type of publication?
\item Which authors and institutions publish the most on the topic?
\item What domains of cybersecurity have already benefited from XAI research?
\item Why is security analysts' ability to interpret AI cyber risk classification important?
\item How are techniques evaluated?
\item What are the limitations of current techniques?
\end{enumerate}

The repositories of scientific articles in which to conduct the search were defined based on the prevalence of use by researchers of information technologies, in addition to allowing access via the Web and search queries:
\begin{enumerate}
\item Scopus (http://www.scopus.com/home.url)
\item ACM Digital Library (http://portal.acm.org/)
\item IEEE Xplore Digital Library (http://ieeexplore.ieee.org/)
\end{enumerate}

Only articles written in \textbf{English} were considered in the scope of this work. In addition, books and panels were disregarded.

A set of keywords was generated, among them the more general: \emph{"explainable artificial intelligence"} and \emph{"cybersecurity"}, and its variation \emph{"cyber security"} (the forms \emph{"cyber-security"} and \emph{"interpretable artificial intelligence"} did not add new results).
Aiming to increase the number of results returned, keywords were added referring to specific cybersecurity topics, so that articles that do not mention "cybersecurity" but that otherwise belong to the area can be included. The added keywords are: 
\begin{enumerate}
\item \emph{"detection and response"}
\item \emph{"intrusion detection"}
\item \emph{"intrusion prevention"}
\item \emph{"cyber risk"}
\item \emph{"malware"}
\end{enumerate}

Thus, the formulated search string is:

\begin{center}
\emph{("explainable artificial intelligence") AND ("cybersecurity" OR "cyber security" OR "detection and response" OR "intrusion detection" OR "intrusion prevention" OR "cyber risk" OR "malware")}
\end{center}

As inclusion criteria for an article returned by the search to be considered relevant for reading and analysis, the following were established:
\begin{enumerate}
\item Is it a primary work, as opposed to other literature reviews?
\item Does the XAI technique discussed have a cybersecurity domain as its main application?
\end{enumerate}

Since the returned articles were in an accessible quantity, there was no need to establish exclusion criteria.

Once all these SLR parameters were defined, the search itself was carried out:
\begin{enumerate}
\item The queries were performed in the established repositories
\item Redundant articles were filtered
\item Titles and abstracts were read, considering the inclusion/exclusion criteria
\item The remaining articles have been read in their entirety
\end{enumerate}

\section{SLR Phase 2: Analysis}
In this phase of the present work, we seek to answer some of the secondary questions, extracting quantitative insights from the results obtained.

During the search phase, 42 valid and unique references were retrieved. Books and panels were considered invalid references. After applying the inclusion/exclusion criteria, 21 papers were considered relevant. More details can be seen in Table~\ref{tab:RecoveredPapers}. One of the retrieved works was another systematic review \cite{Hariharan2021}, whose references were also included in the scope of this work, totaling 36 relevant papers, by the inclusion/exclusion criteria.

In order to answer secondary question number 1, the analysis of which countries publish the most on the subject was carried out from two perspectives. In Table~\ref{tab:PapersByCountry} we associate the paper with the country of the institution to which the first author is linked. Otherwise, in Table~\ref{tab:AuthorsByCountry} we associate each author with the country of the institution to which they are linked (link with more than one country allowed).

To answer secondary question number 2, Table~\ref{tab:PapersByYear} was sorted by year, showing the number of studies on the topic in each year. Furthermore, to shed light on secondary question number 3, Table~\ref{tab:PapersByPubType} shows the number of studies depending on the type of publication. 

Aiming at secondary question number 4, Table~\ref{tab:PapersByAuthor} shows the number of publications by authors with at least two articles. Of the 141 different authors, only 5 published more than one paper within the scope of this work. Also, Table~\ref{tab:PapersByInstitution} shows the number of papers from Institutions that published more than one article. Of the 60 institutions that published a paper, 6 were the ones that published at least two.

\begin{table}[]
\centering
\caption{Papers Retrieved from Queries}
\label{tab:RecoveredPapers}
\begin{tabular}{|l|l|}
\hline
\multicolumn{1}{|c|}{\textbf{Repository}} & \multicolumn{1}{c|}{\textbf{Number of Papers}} \\ \hline
ACM                               & 15          \\ \hline
IEEE                              & 13          \\ \hline
Scopus                            & 24          \\ \hline
\textbf{Total}                    & \textbf{52} \\ \hline\hline\hline
\textbf{Unique and Valid References}         & \textbf{42} \\ \hline
\textbf{After Inclusion Criteria} & \textbf{21} \\ \hline
\end{tabular}
\end{table}
\begin{table}[]
\centering
\caption{Papers per Country}
\label{tab:PapersByCountry}
\begin{tabular}{|l|l|}
\hline
\multicolumn{1}{|c|}{\textbf{Country}} & \multicolumn{1}{c|}{\textbf{Number of Papers}} \\ \hline
USA         & 13 \\ \hline
China       & 3  \\ \hline
India       & 3  \\ \hline
Italy       & 3  \\ \hline
Germany     & 2  \\ \hline
South Korea & 2  \\ \hline
Austria     & 1  \\ \hline
Canada      & 1  \\ \hline
Ireland     & 1  \\ \hline
Israel      & 1  \\ \hline
Japan       & 1  \\ \hline
Mexico      & 1  \\ \hline
Poland      & 1  \\ \hline
Qatar       & 1  \\ \hline
UAE         & 1  \\ \hline
UK          & 1  \\ \hline
\textbf{Total}                         & \textbf{36}                                    \\ \hline
\end{tabular}
\end{table}
\begin{table}[]
\centering
\caption{Authors per Country}
\label{tab:AuthorsByCountry}
\begin{tabular}{|l|l|}
\hline
\multicolumn{1}{|c|}{\textbf{Country}} & \multicolumn{1}{c|}{\textbf{Number of Authors}} \\ \hline
USA           & 52 \\ \hline
Italy         & 21 \\ \hline
China         & 18 \\ \hline
South   Korea & 7  \\ \hline
Germany       & 6  \\ \hline
India         & 4  \\ \hline
Israel        & 4  \\ \hline
Japan         & 4  \\ \hline
Poland        & 4  \\ \hline
Austria       & 3  \\ \hline
Ireland       & 3  \\ \hline
Mexico        & 3  \\ \hline
UK            & 3  \\ \hline
Canada        & 2  \\ \hline
Qatar         & 2  \\ \hline
Spain         & 2  \\ \hline
UAE           & 2  \\ \hline
Czech         & 1  \\ \hline
France        & 1  \\ \hline
Indonesia     & 1  \\ \hline
Yemen         & 1  \\ \hline
\end{tabular}
\end{table}


\begin{table}[]
\centering
\caption{Papers per Year}
\label{tab:PapersByYear}
\begin{tabular}{|l|l|}
\hline
\multicolumn{1}{|c|}{\textbf{Year}} & \multicolumn{1}{c|}{\textbf{Number of Papers}} \\ \hline
2022           & 2           \\ \hline
2021           & 16          \\ \hline
2020           & 15          \\ \hline
2019           & 1           \\ \hline
2018           & 2           \\ \hline
\textbf{Total} & \textbf{36} \\ \hline
\end{tabular}
\end{table}

\begin{table}[]
\centering
\caption{Papers per Publication Type}
\label{tab:PapersByPubType}
\begin{tabular}{|l|l|}
\hline
\multicolumn{1}{|c|}{\textbf{Publication Type}} & \multicolumn{1}{c|}{\textbf{Number of Papers}} \\ \hline
Journal Article        & 26          \\ \hline
Conference Proceedings & 7           \\ \hline
Report                 & 3           \\ \hline
\textbf{Total}         & \textbf{36} \\ \hline
\end{tabular}
\end{table}

\begin{table}[]
\centering
\caption{Papers per Author}
\label{tab:PapersByAuthor}
\begin{tabular}{|l|l|}
\hline
\multicolumn{1}{|c|}{\textbf{Year}} & \multicolumn{1}{c|}{\textbf{Number of Papers}} \\ \hline
Islam, Sheikh Rabiul & 3 \\ \hline
Drichel,   Arthur    & 2 \\ \hline
Eberle, William      & 2 \\ \hline
Mane,   Shraddha     & 2 \\ \hline
Rao, Dattaraj        & 2 \\ \hline
\end{tabular}
\end{table}

\begin{table}[]
\centering
\caption{Papers per Institution}
\label{tab:PapersByInstitution}
\begin{tabular}{|l|l|}
\hline
\multicolumn{1}{|c|}{\textbf{Institution}} & \multicolumn{1}{c|}{\textbf{Number of Papers}} \\ \hline
Chinese Academy of Sciences        & 2 \\ \hline
Persistent Systems Limited         & 2 \\ \hline
RWTH Aachen University             & 2 \\ \hline
Tennessee Technological University & 2 \\ \hline
University of California           & 2 \\ \hline
University of Hartford             & 2 \\ \hline
\end{tabular}
\end{table}

\section{SLR Phase 3: Conclusion}
In this phase we explore the XAI techniques that were applied by the reviewed papers to different areas of cybersecurity. Furthermore, addressing secondary questions 6 to 8, we will note the importance given to explainability, ways of evaluating it, and limitations found against it. A more detailed summary of all texts can be found in Table~\ref{tab:Summary} (Appendix). In it, the authors' motivation for the use of XAI is transcribed, as well as, when mentioned in the articles, the limitations encountered when seeking explainability and the techniques applied to evaluate it.

In Reyes et al. \cite{Reyes2020}, SHapley Additive exPlanations (SHAP) technique is used to understand the influence of features on each type of network traffic records over a machine-learning based IDS.

\cite{Szczepanski2020}'s Intrusion Detection System uses a hybrid approach to deliver maximum accuracy, and still provide explainability. Consists of a Feed Forward Artificial Neural Network (ANN) black-box classifier, named Oracle, and a surrogate explanation module, which is composed of Decision Trees trained using microaggregation. It is model agnostic with local scope explanations.

\cite{Marino2018} takes an adversarial approach to generate explanations for incorrect
classifications made by IDS. They use it to find the minimum modifications of the input features required to correctly classify a given set of misclassified samples.

\cite{Wang2020} develops a framework using SHapley Additive exPlanations (SHAP) to provide local and global explanations on the functioning of an IDS.

\cite{Nellaivadivelu2020} performs an example-based black box analysis of Android anti-malware solutions, to determine which features a detector relies on for its classifications.

\cite{Veksler2020} is particularly innovative in utilizing the intrinsically interpretable Symbolic Deep Learning (SDL) method, which constructs cognitive models based on small samples of expert classifications, to provide decision support for non-expert users in the form of explainable suggestions over Intrusion Detection. Human experiment results reveal that SDL can help to reduce missed threats by 25\%. 

\cite{Kim2021} strongly talks about the need to apply XAI to cybersecurity technologies to improve the efficiency of analysts' decision making. In the paper, SHAP and FOS (feature outlier score) techniques are applied to find valuable information in IDS and Malware datasets.

In \cite{Karn2021}, SHAP, LIME, and an auto-encoding-based scheme for LSTM (Long short-term memory) models are applied to an ML-based detection system for cryptomining in a Kubernetes cluster. 

In \cite{Mahdavifar2020}, a DeNNeS (deep embedded neural network expert system) which extracts refined rules from a trained DNN (deep neural network) to substitute the knowledge base of an expert system is proposed. It's then applied to Phishing Detection and Malware Classification.

In \cite{Melis2020}, gradient-based attribution methods are used to explain Android malware classifiers’ decisions by identifying the most relevant features. Also, the authors propose metrics to evaluate the impact of the explanation on the adversarial robustness of the classifiers.

In \cite{Islam2020}, ML models are infused with Domain Knowledge for Intrusion Detection. They use six different algorithms for predicting malicious records: a probabilistic classifier based on the Naive Bayes theorem, and five supervised “black box” models. Their finding is that "domain knowledge infusion provides better explainability with negligible compromises in performance".

\cite{Holder2021} presents a use case for understanding both what information requirements a human needs for decision-making, as well as what information can be made available by the AI, seeking a guide for the development of future explainable systems. In this particular use case, the XAI takes the role of a junior cyber analyst.

In \cite{Mahbooba2021}, a DT (Decision Tree) model is used for a Intrusion Detection System. The authors point out that previous works that have used Decision Trees in IDS focused on the accuracy of benchmark machine learning algorithms. Conversely, this paper focus on the interpretability of a widely used benchmark dataset.

\cite{Mathews2019} applies LIME (Local Interpretable Model-Agnostic Explanations) to Malware Classification. Also, it has a great discussion on XAI in general, citing important concepts for the evaluation of explainability developed by others, such as Descriptive Accuracy.

In \cite{Antwarg2020}, SHAP is used do explain autoencoder anomaly detections. One of the datasets is of intrusions simulated in a military network environment.

\cite{Ahn2020} uses a GA (Genetic Algorithm) to promote explanations for a network traffic classifier, which in turn can be used for Intrusion Detection. The GA selects important features in the entire feature set.

In \cite{Shraddha2021}, a profusion of XAI techniques is used to improve the interpretability of an Intrusion Detection System. They leverage SHAP, LIME, and three other algorithms present in the AIX360 (AI explainability 360) open-source toolkit by IBM to create a framework that "provides explanations at every stage of machine learning pipeline".

\cite{Liu2021} is a beautiful work on applying XAI to many stages of the human-AI interaction and evaluating it. The authors propose a framework named FAIXID for improving the explainability and understandability of intrusion detection alerts. Their method has been implemented and evaluated using experiments with real-world datasets and human subjects.

In \cite{Wu2020}, a visual analytics system for CNN (Convolutional Neural Network) interpretation, using LIME and Saliency Maps, is described, which is applied in the context of Intrusion Detection. 

In \cite{Drichel2021}, Random Forest, a feature-based machine learning model, is used to identify DGA (Domain-Generation Algorithms), based on Domain Names, in the context of Intrusion Detection.

In \cite{Gu2021}, SHAP is applied to a DNN (Deep Neural Network) in the context of host-based Intrusion Detection, with the specific purpose of better understanding the features to improve the algorithm execution time.

In \cite{Islam2021}, the authors extend their previous work presented in \cite{Islam2020}, now also applying a proxy task-based explainability quantification method.

In \cite{Andresini2021}, addressing the issue of \emph{concept drift} in the context of network Intrusion Detection, the authors propose a framework named INSOMNIA, which, among other functionalities, makes use of DALEX (moDel Agnostic Language for Exploration and eXplanation), an open-source XAI package for Python and R, to understand feature importance changes over time.

In \cite{Becker2020}, a visual analytics system is applied to interpret two different types of deep learning-based neural nets for Domain-Generation Algorithm (DGA) classification, in the context of Intrusion Detection. It works by clustering the activations of a model’s neurons and subsequently leveraging decision trees in order
to explain the constructed clusters. In combination with a 2D projection, the user can explore how the model views the data at different layers.

In \cite{Guo2018}, LEMNA (Local Explanation Method using Nonlinear Approximation) a novel, high-fidelity explanation method dedicated for security applications is developed. In the paper, it is used with two deep learning applications in security: Malware Classification, and Binary Reverse-Engineering. The authors also care about demonstrating the practical applications of the explanation method.

In \cite{Song2020}, an RL (Reinforcement Learning) Adversarial approach is taken to evade PE (Portable Executable) Malware classifiers. It is able to shed light on the root cause of the evasions and thus provide feature interpretation.

In \cite{Suryotrisongko2022}, four XAI techniques and Open-Source Intelligence (OSINT) are blended  to deliver better AI explainability through second opinion approaches. The techniques are ANCHOR, LIME, SHAP, and Counterfactual Explanations and they are applied to a Domain-Generation Algorithm (DGA) classifier, for Intrusion Detection.

In \cite{Aguilar2022}, a decision-tree-based autoencoder is described, designed to detect anomalies and provide the explanations behind its decisions by finding the correlations among different attribute values. 

In \cite{Iadarola2021}, applies heatmaps generated with Grad-CAM to interpret a deep-learning based mobile malware classifier.

In \cite{Zolanvari2021}, the authors propose a universal XAI model named Transparency Relying Upon Statistical Theory (TRUST), which is model-agnostic, high-performing, and suitable for numerical applications. They demonstrate the effectiveness of TRUST in a case study on the Industrial Internet of things (IIoT) using three different datasets.

In \cite{Feichtner2020}, the authors propose a system to shed light on how an app description reflects privacy-related permission usage, in the context of Mobile Apps. They apply LIME to their CNN (Convolutional Neural Network) in order to assess the quality of their network and to avoid incomprehensible black box predictions.

\cite{Nascita2021} uses SHAP with a Multimodal DL-based Mobile Traffic Classifier to evaluate the input importance.

In \cite{Roshan2021}, SHAP is used to explain autoencoder anomaly detections. 

In \cite{Aghakhani2020}, aiming to answer the question "\emph{does static analysis on packed binaries provide rich enough features to a malware classifier?}", the authors use a Random Forest method with feature selection.

In \cite{Gulmezoglu2021}, LIME and Saliency Maps are applied successfully to black-box models that are used for WF (Website Fingerprinting) attacks, to explore the leakage sources. The authors also evaluate the usage of the techniques with the Remove and Retrain (ROAR) metric for explainability.

Finally, \cite{Dattaraj2021} addresses the Alarm Flooding problem so common to Intrusion Detection and SIEM (Security Information and Event Management) systems, by automatically labeling the alerts and categorizing them. To this end, the authors use a ZSL (Zero-shot Learning) method interpreted through SHAP and LIME.
\section{Concluding Remarks}
Cybersecurity is a growing concern for businesses and governments. Vendors of cybersecurity solutions are increasingly using AI (Artificial Intelligence). The ability to explain the decisions of an AI algorithm brings several benefits, including greater confidence in the system and a better understanding of its operation. In this sense, XAI (eXplainable Artificial Intelligence) is being applied in several areas of cybersecurity. 

In this systematic literature review, we sought to discover which XAI techniques have been applied in cybersecurity, and which areas of cybersecurity have already benefited from this technology.

Almost all of the works reviewed in this paper explicitly mention some reason why explainability is important. Interestingly, the reasons varied a lot between, on the one hand, improving users' trust in the system, and, on the other hand, enabling researchers to understand the internal mechanisms of the classifier.

As we can see in Table~\ref{tab:Summary}, few papers perform tests that specifically assess the degree of explainability of the technique employed or its practical impact. In this sense, reading \cite{Veksler2020} and \cite{Liu2021} is particularly recommended, due to the excellent description of the techniques used and for carrying out human experiments in the evaluation of the practical consequences of explainability. \cite{Gulmezoglu2021} is a good example of using the ROAR (Remove and Retrain) explainability metric.

Also referring to Table~\ref{tab:Summary}, a minority of texts pointed out the limitations caused by adding more explainability to their algorithms. Among the XAI limitations pointed out are the decrease in the accuracy of intrinsically explainable models, performance difficulties, and the lack of formalization in the concept of \emph{explanation}.

We can see that SHAP and LIME techniques are the most used, perhaps because they have been implemented in open-source frameworks for some time. LEMNA \cite{Guo2018} appears to be a promising technique, developed with cybersecurity use cases in mind. 

Intrusion Detection, Malware Classification, Phishing Detection, Reverse Engineering, Website Fingerprinting, Domain-Generation Algorithms Detection and Abuse of Privacy-related Permissions on Mobile Apps are areas of cybersecurity that have already made use of XAI.

The authors hope that this work can encourage and contribute to the adoption of XAI in more areas of cybersecurity.
\bibliographystyle{elsarticle-num} 
\bibliography{biblio}

\section{Appendix}
The following appendix contains a more detailed summary of all the papers reviewed in this work.
\clearpage
\onecolumn
\newcolumntype{P}[1]{>{\raggedright\let\newline\\\arraybackslash\hspace{0pt}}p{#1}}
\begin{landscape}
\begin{longtable}{|P{0.05\textwidth}|P{0.15\textwidth}|P{0.15\textwidth}|P{0.32\textwidth}|P{0.25\textwidth}|P{0.25\textwidth}|}
\caption{Summary of Papers}
\label{tab:Summary}\\
\toprule
\textbf{Paper} & \textbf{Technique} & \textbf{CyberSec Area} & \textbf{XAI Importance} & \textbf{Evaluation of Explainability} & \textbf{Limitations} \\* \midrule
\endhead
\cite{Reyes2020} 
  & SHAP (SHapley Additive exPlanations)
  & Intrusion Detection 
  & "\emph{XAI was implemented to have an insight for the decisions made by the first stage ML model, mostly for the cases where the records were predicted as impersonation or injection. The features that significantly contribute to their prediction were determined.}" 
  & "\emph{}"
  & "\emph{}"
   \\* \midrule
   
\cite{Szczepanski2020}
  & Decision Trees with Microaggregation
  & Intrusion Detection
  & "\emph{The ability to understand how a system makes a decision is necessary to help develop trust, settle issues of fairness and perform the debugging of a model.}"
  & "\emph{}"
  & "\emph{The derived explanation is not a faithful representation of the opaque classifier function in general}", referring to the \emph{fidelity} feature in AI desiderata \cite{Hansen2019}.
   \\* \midrule

\cite{Marino2018}
  & Example based, Adversarial
  & Intrusion Detection
  & "\emph{It is crucial that the inner workings of data-driven models are transparent for the engineers designing IDSs. Decisions presented by explainable models can be easily interpreted by a human, simplifying the process of knowledge discovery. Explainable approaches help on diagnosing, debugging, and understanding the decisions made by the model, ultimately increasing the trust on the data-driven IDS.}"
  & "\emph{}"
  & "\emph{}"
   \\* \midrule

\cite{Wang2020}
  & SHAP
  & Intrusion Detection
  & "\emph{It is imperative to provide some information about the reasons behind IDSs predictions, and provide cybersecurity personnel with some explanations about the detected intrusions}" and also "\emph{This framework contributes to a deeper understanding of the predictions made from IDSs, and ultimately help build cyber users’ trust in the IDSs.}"
  & "\emph{}"
  & "\emph{}"
   \\* \midrule

\cite{Nellaivadivelu2020}
  & Example based, Adversarial
  & Malware Classification
  & "\emph{Such analysis will help us to understand the robustness of detectors when dealing with minor variants of known malware samples. The second issue concerns the possibility of uncovering important aspects of a malware detection algorithm. Thus, black box analysis of malware detectors can point towards ways to improve on existing malware detectors}"
  & "\emph{}"
  & "\emph{}"
   \\* \midrule

\cite{Veksler2020}
  & SDL (Symbolic Deep Learning)
  & Intrusion Detection
  & "\emph{We project that SDL-generated cognitive models of expert analysts will impart a high degree of trust for at least two reasons – (...) SDL promises to be a more transparent technique than DL, one that is able to provide some explainability for each of its suggestions}"
  & "\emph{Human experiment results reveal that SDL can help to reduce missed threats by 25\%.}"
  & "\emph{The major hurdle for symbolic deep models of memory has been a combinatoric explosion of memory.}"
   \\* \midrule

\cite{Kim2021}
  & SHAP, FOS (Feature Outlier Score)
  & Intrusion Detection, Malware Classification
  & "\emph{AI for cyber security requires final confirmation by an analyst, e.g. malware misdetection can cause significant adverse side effects. Thus, a human analyst must check all AI predictions, which poses a major obstacle to AI expansion. [XAI] enable analysts with limited daily workload to focus upon valuable data, and quickly verify AI predictions.}"
  & "\emph{}"
  & "\emph{}"
   \\* \midrule

\cite{Karn2021}
  & SHAP, LIME (Local Interpretable Model-Agnostic Explanations), and an auto-encoding-based scheme for LSTM (Long Short-Term Memory) models
  & Malware Classification
  & "\emph{The explanation will justify and support disruptive administrative decisions}" and to answer the questions "\emph{Why did the ML classify a particular pod as a miner? How does the syscalls sequence change from one pod to another? Which feature has the greatest impact on miner prediction? Is there any way to visualize the ML outcome apart from plotting the evaluation metrics?}"
  & "\emph{The performance of an autoencoder model is evaluated based on the model’s ability to recreate the input sequence. Validation of the autoencoder model also validates the upstream half of the classifier model which, in turn, further strengthens the trust in the classifier’s outcome.}"
  & "\emph{Convergence is a major issue with autoencoder design, especially when the dataset size is large, and the centroids of the various classes have significant variance. Also, when convergence is achieved, it is often the case that it is at a local minimum of the loss function. Such difficulties impact the quality of the explainability method.}"
   \\* \midrule

\cite{Mahdavifar2020}
  & Deep Embedded Neural Network Expert System 
  & Malware Classification, Phishing Detection
  & "\emph{Security experts not only do need to detect the incoming threat but also need to know the incorporating features that cause that particular security incident}" and "\emph{Adding an explanation feature to a neural network would enhance its trustworthiness and reliability.}"
  & "\emph{}"
  & "\emph{}"
   \\* \midrule

\cite{Melis2020}
  & Gradient-based Explanations
  & Malware Classification
  & "\emph{We investigate whether gradient-based attribution methods used to explain classifiers’ decisions provide useful information about the robustness of Android malware detectors against sparse attacks.}"
  & "\emph{We propose and empirically validate a few synthetic metrics that allow correlating the evenness of gradient-based explanations with the classifier robustness to adversarial attacks.}"
  & "\emph{}"
   \\* \midrule

\cite{Islam2020}
  & Domain Knowledge Infusion
  & Intrusion Detection
  & "\emph{The lack of explainability and interpretability of successful AI models is a key stumbling block when trust in a model’s prediction is critical. This leads to human intervention, which in turn results in a delayed response or decision}"
  & The authors conduct an Explainability Test whose purpose "\emph{is to discover the comparative advantages or disadvantages of incorporating domain knowledge in the experiment}"
  & The authors stress that "\emph{there are some open challenges surrounding explainability and interpretability such as an agreement of what an explanation is and to whom, a formalism for the explanation, and quantifying the human comprehensibility of the explanation}"
   \\* \midrule

\cite{Holder2021}
  & -
  & Cybersecurity Operations
  & "\emph{There are many applications where artificial intelligence (AI) can add a benefit, but this benefit may not be fully realized, if the human cannot understand and interact with the output as required by their context. Allowing AI to explain its decisions can potentially mitigate this issue.}"
  & "\emph{}"
  & "\emph{}"
   \\* \midrule

\cite{Mahbooba2021}
  & Decision Tree
  & Intrusion Prevention
  & "\emph{eXplainable Artificial Intelligence (XAI) has become increasingly important to interpret the machine learning models to enhance trust management by allowing human experts to understand the underlying data evidence and causal reasoning}"
  & "\emph{}"
  & "\emph{There may be a chance of overfitting when the algorithm captures noise in the dataset}", besides that "\emph{Information gain in decision trees is biased in favor of those attributes with more levels. This behavior might impact prediction performance.}"
   \\* \midrule

\cite{Mathews2019}
  & LIME
  & Malware Classification
  & "\emph{It enables human users to understand, appropriately trust, and effectively manage the emerging generation of artificially intelligent partners}" and also "\emph{Explainability in general also helps to identify bias in raw data and strategize   the model optimization}"
  & The paper defines the concept of "\emph{descriptive accuracy}" as the ability of the interpretations to properly describe what the model has learned. Although it mentions "descriptive accuracy", it does not evaluate it against the applied technique. Nonetheless, the system is evaluated through use cases, showing an analysis of why a model makes mistakes.
  & LIME technique does not produce “fixed” feature importance plots (i.e., a general rather than a case-to-case view of which variables are most informative when making a prediction). The  explanation  reflects the behavior of the classifier “around” the instance being predicted.
   \\* \midrule

\cite{Antwarg2020}
  & SHAP
  & Anomaly Detection, Intrusion Detection
  & "\emph{The manual validation of results becomes challenging without justification or additional clues. An explanation of why an instance is anomalous enables the experts to focus their investigation on the most important anomalies and may increase their trust in the algorithm}"
  & "\emph{}"
  & "\emph{}"
   \\* \midrule

\cite{Ahn2020}
  & Feature Selection with Genetic Algorithm
  & Intrusion Detection
  & "\emph{The mechanism of deep learning is inexplicable. A malfunction of the deep learning model may occur if the training dataset includes malicious or erroneous data. Explainable artificial intelligence (XAI) can give some insight for improving the deep learning model by explaining the cause of the malfunction}"
  & "\emph{}"
  & "\emph{}"
   \\* \midrule

\cite{Shraddha2021}
  &  SHAP, LIME, Contrastive Explanations Method (CEM), ProtoDash and Boolean Decision Rules via Column Generation (BRCG)
  & Intrusion Detection
  & "\emph{Deep neural networks are complex and hard to interpret which makes difficult to use them in production as reasons behind their decisions are unknown.}" and also "\emph{We propose an explainable AI framework along with intrusion detection system which would help analyst to make final decision.}"
  & "\emph{}"
  & "\emph{}"
   \\* \midrule

\cite{Liu2021}
  & Boolean Rule Column Generation (BRCG), Logistic Rule Regression(LogRR), ProtoDash, Contrastive Explanations Method (CEM)
  & Intrusion Detection
  & "\emph{The decisions from AI solutions have to be explainable to gain analysts’ trust and help analysts in making a confident and accountable decision.}" and also "\emph{We need XAI to improve fairness, accountability, and trust in decisions}"
  & The researchers conducted extensive evaluation of explainability using human subject and proxy methods experiments.
  & "\emph{}"
   \\* \midrule

\cite{Wu2020}
  & LIME and Saliency Maps 
  & Intrusion Detection
  & "\emph{Researchers in the field of network security are the target users of our visual analytics system. The design goal of our visual analytics system is to aid our target users to better interpret the deep learning model.}"
  & "\emph{}"
  & No case study on real datasets; lack of scalability for larger DL models; limited to CNNs.
   \\* \midrule

\cite{Drichel2021}
  & Random Forest
  & DGA Classification, Intrusion Detection
  & "\emph{The proposed state-of-the-art classifiers are based on deep learning models. The black box nature of these makes it difficult to evaluate their reasoning. The resulting lack of confidence makes the utilization of such models impracticable}"
  & "\emph{}"
  & "\emph{}"
   \\* \midrule

\cite{Gu2021}
  & SHAP
  & Intrusion Detection
  & "\emph{Help us understand how deep learning models learn and why they make such decisions for each input}" and also "\emph{We propose a method to improve detection efficiency by using XAI to reduce the input data}"
  & "\emph{}"
  & "\emph{}"
   \\* \midrule

\cite{Islam2021}
  & Domain Knowledge Infusion
  & Intrusion Detection
  & "\emph{The lack of explainability leads to a lack of trust in the model and prediction, which can involve ethical and legal issues in critical domains due to the potential implications on human interests, rights, and lives}"
  & The authors extend the work made in \cite{Islam2020}, applying to it the Explainability Quantification Method previously developed by them in \cite{Islam2019}.
  & "\emph{}"
   \\* \midrule

\cite{Andresini2021}
  & DALEX (moDel Agnostic Language for Exploration and eXplanation)
  & Intrusion Detection
  & "\emph{Apply explainable AI to better interpret how the model reacts to the shifting distribution.}"
  & "\emph{}"
  & "\emph{}"
   \\* \midrule

\cite{Becker2020}
  & Decision Tree, Visual Analytics
  & DGA Classification, Intrusion Detection
  & "\emph{Deep learning models
have found wide adoption for many problems. However, their blackbox nature makes it hard to trust their decisions and to evaluate their line of reasoning. In the field of cybersecurity, this lack of trust and understanding poses a significant challenge for the utilization of deep learning models}"
  & "\emph{}"
  & "\emph{}"
   \\* \midrule

\cite{Guo2018}
  & LEMNA (Local Explanation Method using Nonlinear Approximation)
  & Malware Classification, Binary Reverse Engineering
  & "\emph{Security practitioners are concerned about the lack of transparency of the deep learning models and thus hesitated to widely adopt deep learning classifiers in security and safety-critical areas}"
  & "\emph{The fidelity metrics are computed either by directly comparing the approximated detection boundary with the real one, or running end-to-end feature tests}"
  & "\emph{}"
   \\* \midrule

\cite{Song2020}
  & Adversarial
  & Malware Classification
  & "\emph{Researchers should use explanation techniques to understand the behavior of the classifiers and check if the learned features are fragile features that can be easily evaded or if they conflict with expert knowledge}"
  & "\emph{}"
  & "\emph{}"
   \\* \midrule

\cite{Suryotrisongko2022}
  & ANCHOR, LIME, SHAP and Counterfactual Explanations
  & DGA Classification, Intrusion Detection
  & XAI and Open-Source Intelligence can together address "\emph{trust problems}", serving as "\emph{an antidote for skepticism to the shared models and preventing automation bias.}"
  & "\emph{}"
  & "\emph{}"
   \\* \midrule

\cite{Aguilar2022}
  & Decision Trees
  & Anomaly Detection
  & They cite other authors stating that "\emph{explanations ensure the correct behavior of the algorithm}" and "\emph{machine learning systems would be more widely accepted once they are capable of providing satisfactory explanations for their decisions.}"
  & "\emph{}"
  & "\emph{First, our architecture should be used in datasets with less than a thousand attributes because it builds a tree for each attribute. Building thousands of trees is time-consuming, although this limitation may be overcome with access to better computing resources. Second, our proposal may fail to build decision trees for attributes with tens of different values in the definition domain.}"
   \\* \midrule

\cite{Iadarola2021}
  & Heatmaps
  & Malware Classification
  & "\emph{The XAI aims to enable human users to develop understanding and trusts to the model prediction.}" and also "\emph{The effectiveness of these [autonomous] systems is limited by the current inability of machines to explain their decisions and actions to human users. The most important step towards reliable models is the possibility to understand their prediction i.e., the so-called interpretability.}"
  & "\emph{}"
  & "\emph{Despite their usefulness, the cumulative heatmaps, at the time of writing, do not play a role that can be automatized without knowledge on the dataset and the malware code. They help to interpret and understand the outcomes, but they do not provide fixed information that could be used to any user to evaluate models without any prior-knowledge on the architecture or the problem itself.}"
   \\* \midrule

\cite{Zolanvari2021}
  & TRUST (Transparency Relying Upon Statistical Theory)
  & Intrusion Detection
  & "\emph{Despite the popularity of AI, it is limited by its current inability to build trust. Researchers and industrial leaders have a hard time explaining the decisions that sophisticated AI algorithms come up with because they (as AI users) cannot fully understand why and how these “black boxes” make their decisions.}"
  & "\emph{}"
  & "\emph{Due to using information gain in picking the representatives, TRUST might overfit to the training set. This would lead to poor performance on unseen data. On the other hand, if the Gaussian assumption cannot be made or the probability distribution of data changes, the output of TRUST would not be reliable. Also, the assumption of samples being drawn independently is very important.}"
   \\* \midrule

\cite{Feichtner2020}
  & LIME
  & Abuse of Privacy-related Permissions on Mobile Apps
  & "\emph{To assess the quality of our network and to avoid incomprehensible black box predictions, we employ the model explaining algorithm LIME}"
  & "\emph{}"
  & "\emph{}"
   \\* \midrule

\cite{Nascita2021}
  & Deep SHAP
  & Network Traffic Classifier, Intrusion Detection
  & "\emph{The black-box nature of DL techniques hides the reason behind specific classification outcomes. This impacts the understanding of classification errors and the evaluation of the resilience against adversarial manipulation of traffic to impair identification. Moreover, by understanding the behavior of the learned model, performance enhancements can be pursued with much more focused and efficient research, compared with a less-informed exploration of the (typically huge) hyper-parameters space.}"
  & "\emph{}"
  & "\emph{}"
   \\* \midrule

\cite{Roshan2021}
  & SHAP
  & Anomaly Detection, Intrusion Detection
  & "\emph{In making life-changing decisions such as disease diagnosis, it's crucial to understand why the system makes such a critical decision. Hence the importance of explaining the AI system. Furthermore, the black-box nature of the AI-based system gives excellent results but without any explanation, and hence, they lose their trust to adapt these systems in critical decision making.}"
  & "\emph{}"
  & "\emph{}"
   \\* \midrule

\cite{Aghakhani2020}
  & Random Forest
  & Malware Classification
  & The authors only discuss the results of the Random Forest approach in their paper because "\emph{Random forest allows for better interpretation of the results compared to neural networks}".
  & "\emph{}"
  & "\emph{}"
   \\* \midrule

\cite{Gulmezoglu2021}
  & LIME and Saliency Maps
  & Website Fingerprinting Detection
  & "\emph{The lack of XAI studies on Website Fingerprinting slows down the research on countermeasures against this type of attack since the leakage source is not clearly visible to both attackers and cyber-defenders. Therefore, there is a need for a sophisticated analysis technique to identify the leakage sources in the side-channel data by applying XAI algorithms to trained models.}"
  & "\emph{ROAR metric is implemented on both techniques and it is shown that LIME and saliency map correctly discover the most dominant features in the side-channel measurements.}"
  & "\emph{In this study, LIME cannot be applied on CNN model due to the lack of high performance.}"
   \\* \midrule

\cite{Dattaraj2021}
  & SHAP, LIME
  & Alarm Flooding, Intrusion Detection
  & "\emph{Explanations give us measurable factors as to what features influence the prediction of a cyber-attack and to what degree}" and also "\emph{Without any prior knowledge of the attack, we try to identify it, decipher the features that contribute to its classification and try to bucketize the attack in a specific category - using explainable AI}"
  & "\emph{}"
  & "\emph{}"
   \\* \bottomrule
\end{longtable}
\end{landscape}
\clearpage
\twocolumn
\end{document}